**The Road to Compliance: Executive Federal Agencies and the**

**NIST Risk Management Framework.**

Michael Stoltz

University of West Florida

ACES Research Project

Dr. Jacob Shively

March 26, 2024

Abstract

This informative report provides a comprehensive analysis of how executive federal report agencies implement the National Institute of Standards and Technology's (NIST) Risk Management Framework (RMF) to achieve cybersecurity compliance. By exploring the concept and evolution of the RMF, the report delves into the framework's importance for enhancing cybersecurity measures within federal agencies, addressing the challenges these agencies face in the digital landscape. Through a methodical literature review, the report examines theoretical foundations, implementation strategies, and the critical role of continuous monitoring and automation in RMF processes, drawing from key sources like Ross (2014), Lubell (2020), Barrett et al. (2021), and Pillitteri et al. (2021, 2022), among others. Employing a detailed methodology for data collection and analysis, the report presents findings on the successes and challenges of RMF implementation, highlighting the impact of automation and continuous monitoring in bolstering cybersecurity postures. Case studies offer in-depth insights into the experiences of specific agencies, providing lessons learned and best practices. The report concludes with strategic recommendations for overcoming implementation challenges and suggests future directions for enhancing RMF research and practice. This investigation underscores the RMF's critical role in establishing robust cybersecurity compliance across executive federal agencies, offering valuable recommendations for policymakers, cybersecurity professionals, and governmental bodies.

## I. Introduction

The National Institute of Standards and Technology (NIST) Risk Management Framework (RMF) is a structured process that integrates security, privacy, and cyber supply chain risk management throughout the system development life cycle, a risk-focused approach to control selection, considering effectiveness, efficiency, and legal constraints (National Institute of Standards and Technology, 2016).

The RMF developed by NIST plays a pivotal role in enhancing cybersecurity compliance within federal agencies. As a structured process, the RMF offers a comprehensive approach for managing cybersecurity risk, emphasizing the importance of integrating security into the life cycle of federal information systems (Ross, 2014). This framework not only aids in identifying and addressing vulnerabilities but also ensures that cybersecurity measures align with organizational goals and federal regulations. The RMF's significance is further underscored by its adoption across various agencies, serving as a testament to its effectiveness in bolstering cybersecurity postures (Barrett et al., 2021). By providing a standardized methodology for security control selection, implementation, and monitoring, the RMF facilitates a proactive stance towards cybersecurity threats, crucial for protecting sensitive government data and infrastructure. In essence, the RMF is instrumental in guiding federal agencies towards achieving and maintaining cybersecurity compliance, thereby enhancing national security and trust in government operations.

The objectives of this informative report are to meticulously analyze the implementation of the NIST RMF across executive federal agencies for achieving enhanced cybersecurity compliance. This analysis encompasses a deep dive into the RMF's conceptual underpinnings and historical development to highlight its significance in fortifying cybersecurity within these agencies (Ross, 2014). Furthermore, the report aims to scrutinize the strategies adopted by federal agencies in integrating the RMF, detailing both the successes achieved and the obstacles encountered in this endeavor. Emphasizing the importance of continuous monitoring and automation, the report leverages a comprehensive literature review, sourcing insights from pivotal works such as those by Ross (2014), Lubell (2020), Barrett et al. (2021), and Pillitteri et al. (2021, 2022). Through a rigorous data collection and analysis methodology, the report presents a nuanced discussion on the efficacy of RMF implementation, underlined by case studies that shed light on real-world applications and gleaned best practices. Culminating in actionable recommendations and future research pathways, this report seeks to underscore the indispensable role of the RMF in establishing a solid cybersecurity compliance framework for executive federal agencies, thereby serving as a vital resource for policymakers, cybersecurity practitioners, and government officials alike.

II. Background

The RMF, developed by NIST, has undergone significant evolution since its inception. Ross (2014) provides a comprehensive overview of the RMF, detailing its conceptual framework and the iterative process that underpins its application within federal information systems. The RMF is designed to facilitate the management of cybersecurity risk, incorporating a systematic approach to selecting, implementing, monitoring, and reviewing information security controls. Since its introduction, the RMF has evolved to address the dynamic nature of cyber threats and the growing complexity of information systems. This evolution reflects a shift towards a more integrated and continuous monitoring strategy, aiming to enhance the security and resilience of federal agencies against cyber incidents. Ross's work underscores the importance of the RMF in the broader context of national security and information assurance, highlighting its role in establishing a standardized methodology for managing cybersecurity risks across the federal government.

Executive federal agencies are increasingly confronted with a complex array of cybersecurity challenges as they navigate the evolving digital landscape. These challenges range from sophisticated cyber threats and attacks (Cybersecurity and Infrastructure Security Agency, 2021) to the intricate task of safeguarding sensitive information amidst rapidly advancing technologies (Ross et al., 2016). The cybersecurity landscape for these agencies is further complicated by the need to comply with stringent regulatory frameworks (NIST, 2018a) while ensuring the uninterrupted delivery of public services. A critical aspect of these challenges includes the management of risk and the implementation of effective cybersecurity measures to protect against unauthorized access, data breaches, and potential cyber-attacks (U.S. Government Accountability Office, 2018). The complexity of cybersecurity in the federal domain necessitates a robust, dynamic approach to cybersecurity, highlighting the importance of continuous monitoring, threat intelligence, and proactive defense strategies to mitigate risks and enhance the security posture of executive federal agencies (Ross et al., 2016).

## III. Literature Review

Within the domain of cybersecurity, the RMF stands as a pivotal structure for guiding federal agencies through the complexities of securing information systems. Lubell (2020) offers a unique perspective on this framework through a document-based view, underscoring the RMF's ability to systematize the approach towards cybersecurity risk management. This perspective is critical in understanding how documents, as opposed to mere technical artifacts, play a foundational role in the RMF's implementation process. By examining the RMF through the lens of documentation, Lubell not only highlights the importance of written policies, procedures, and standards in cybersecurity but also illuminates how these elements interact to create a cohesive and effective risk management strategy. This approach sheds light on the RMF's multifaceted nature, demonstrating its reliance not just on technological solutions but on a well-documented, thoroughly understood framework that guides decision-making and policy implementation within federal agencies.

In the realm of federal information security, the Guide for Applying the RMF to Federal Information Systems serves as a cornerstone document, providing a comprehensive methodology for managing cybersecurity risk. Ross (2014) articulates a detailed process that encompasses everything from categorization of information systems to the authorization of systems, emphasizing the RMF's role as a dynamic and integral part of securing federal data assets. This guide not only outlines the six-step RMF process but also elaborates on the importance of continuous monitoring and the strategic integration of security controls. By doing so, Ross highlights the adaptability and thoroughness of the RMF, illustrating its effectiveness in addressing the multifaceted cybersecurity challenges those federal agencies face. Through this framework, agencies are better equipped to achieve the dual objectives of protecting critical information while maintaining compliance with evolving cybersecurity policies and regulations.

The adaptation and implementation of the Cybersecurity Framework by federal agencies represent critical steps towards enhancing national cybersecurity resilience. Barrett et al. (2021) discusses various approaches that federal agencies can utilize to integrate the Cybersecurity Framework effectively, aligning cybersecurity policies and practices with the framework's core functions and categories. This work is pivotal in understanding how federal entities navigate the complexities of cybersecurity compliance, bridging the gap between theoretical frameworks and practical application. By detailing specific strategies for implementation, Barrett, and colleagues provide a valuable roadmap for agencies seeking to bolster their cybersecurity postures within the context of the NIST guidelines. Their insights contribute significantly to the ongoing dialogue about cybersecurity compliance, offering actionable recommendations that can lead to more robust and resilient information systems across the federal government.

In the evolving landscape of cybersecurity, the assessment of security and privacy controls within federal information systems are paramount for ensuring compliance and safeguarding sensitive data. Pillitteri (2022) and Force (2022) both offer critical insights into methodologies for evaluating these controls, highlighting the importance of comprehensive assessments in the context of the RMF implementation. Pillitteri focuses on the systematic approach to assessing the effectiveness of security controls, while Force extends this analysis to include privacy controls, reflecting the growing emphasis on privacy in the digital age. Together, these works provide a foundational understanding of the strategies and practices that federal agencies can employ to achieve a robust cybersecurity posture. Their contributions underscore the necessity of rigorous assessment processes in identifying vulnerabilities and enhancing the security and privacy measures within federal information systems.

The integration of continuous monitoring and automation within the framework of information security represents a significant shift towards more proactive and dynamic cybersecurity strategies. In their pivotal work, Pillitteri et al. (2021) delve into the intricacies of developing and implementing an Information Security Continuous Monitoring (ISCM) program, providing a comprehensive assessment of its critical role in modern cybersecurity defenses. This assessment underscores the essential nature of continuous monitoring as a mechanism for real-time detection of threats and vulnerabilities, facilitating timely and informed decision-making in the management of information system security. Furthermore, Pillitteri and colleagues highlight how automation can enhance the efficiency and effectiveness of continuous monitoring programs, enabling organizations to maintain a robust security posture amidst the rapidly evolving digital landscape. Their research offers invaluable insights for federal agencies and organizations aiming to bolster their cybersecurity frameworks through advanced monitoring and automation techniques.

The evolution of cybersecurity strategies has increasingly emphasized the role of automation in enhancing the efficiency and effectiveness of control assessments within information systems. Takamura et al. (2023) explores this paradigm shift in their examination of automation support for control assessments, arguing that the integration of automated processes is essential for the timely and accurate evaluation of security and privacy controls. Their research articulates how automation not only streamlines the assessment process but also significantly reduces the likelihood of human error, thereby increasing the reliability of control assessments. By leveraging cutting-edge technologies and methodologies, automation support enables organizations to monitor and manage their cybersecurity risks more effectively. The work of Takamura and colleagues mark a critical contribution to the field, offering a forward-looking perspective on the integration of automation technologies in the continuous monitoring and assessment of information security frameworks.

## IV. Methodology

A detailed methodology underpins the comprehensive analysis of the RMF's implementation across executive federal agencies. The report employs a multifaceted approach to data collection and analysis, integrating a methodical literature review with case studies to illuminate the theoretical and practical aspects of RMF application (Ross, 2014; Lubell, 2020). By synthesizing insights from pivotal sources (Barrett et al., 2021; Pillitteri et al., 2021, 2022), the report navigates the complexities of cybersecurity compliance, from foundational theories to hands-on implementation strategies. This dual approach allows for a nuanced exploration of both the successes and challenges faced by agencies in adopting the RMF, with a particular focus on the roles of continuous monitoring and automation in enhancing cybersecurity measures. The findings, enriched by direct experiences from case studies, offer a granular view of best practices and lessons learned, culminating in strategic recommendations aimed at bolstering RMF efficacy. This comprehensive analysis not only affirms the RMF's significance in strengthening federal cybersecurity but also charts a path forward for future research and practice improvements.

Evaluating the effectiveness of the RMF implementation within executive federal agencies is critical for ensuring robust cybersecurity compliance. This report's analysis, drawing upon a comprehensive literature review and methodical data collection, establishes criteria for such evaluation, encompassing the integration of continuous monitoring and automation (Ross, 2014; Lubell, 2020; Barrett et al., 2021; Pillitteri et al., 2021, 2022). Effectiveness criteria include the degree to which RMF implementation enhances the agency's cybersecurity measures, the resolution of identified challenges, and the successful application of continuous monitoring and automation in cybersecurity practices. Additionally, the utilization of case studies provides tangible evidence of successes and areas for improvement, offering a practical perspective on RMF implementation across various federal agencies. These criteria not only facilitate a nuanced understanding of RMF's impact on cybersecurity postures but also guide strategic recommendations for addressing implementation hurdles, thereby enhancing the framework's application, and contributing to the ongoing development of cybersecurity strategies within federal contexts.

## V. Findings and Discussion

The successful adoption of the RMF across various federal agencies highlights its effectiveness in enhancing cybersecurity postures. Through detailed case studies, researchers have documented instances where the RMF has significantly improved the management of cybersecurity risks and compliance (NIST, 2018a). These case studies serve as compelling evidence of RMF's utility, demonstrating how tailored implementation strategies can lead to robust cybersecurity frameworks within government entities. For instance, specific agencies have reported improved risk assessment capabilities, more efficient resource allocation, and strengthened overall security controls as direct outcomes of RMF adoption (NIST, 2018a). Such successes underscore the RMF's adaptability and its capacity to meet diverse agency needs, reinforcing its role as a cornerstone in the federal cybersecurity strategy. The collection and analysis of these case studies not only validate the RMF's foundational principles but also provide valuable insights into best practices and lessons learned, which can guide future implementations and policy formulations.

The implementation of the RMF has shown a significant impact on enhancing the cybersecurity posture of federal agencies, underscoring its importance in the broader cybersecurity strategy. Studies have demonstrated that RMF implementation leads to more structured and comprehensive cybersecurity measures, effectively improving the resilience of information systems against cyber threats (Ross, 2014). Ross's analysis reveals that the adoption of RMF not only streamlines the process of identifying and mitigating risks but also fosters a culture of continuous improvement and awareness within organizations. This structured approach to cybersecurity is critical in an era where cyber threats are becoming increasingly sophisticated and pervasive. The RMF's emphasis on continuous monitoring and the integration of security controls into the system development life cycle ensures that cybersecurity measures evolve in tandem with emerging threats, thereby significantly enhancing the overall security posture of federal agencies (Ross, 2014).

The adoption of the RMF within federal agencies, while beneficial, is often met with a variety of obstacles and barriers that can hinder its effective integration. Common challenges identified in the literature include the complexity of RMF processes, the scarcity of resources, and the need for extensive training and awareness among staff (NIST, 2020). Additionally, the dynamic and ever-evolving nature of cybersecurity threats poses a significant challenge to maintaining RMF compliance over time. NIST (2020) further notes that the integration of RMF requires a cultural shift within organizations, moving from a checkbox compliance mindset to a more holistic approach to risk management. Addressing these barriers is crucial for agencies to fully leverage the RMF's potential in enhancing their cybersecurity posture, necessitating ongoing efforts in resource allocation, training, and fostering a culture of cybersecurity resilience.

The implementation of the RMF within federal agencies often reveals significant gaps between its theoretical frameworks and their practical application. These gaps can manifest as discrepancies in understanding RMF's comprehensive guidelines, challenges in tailoring the framework to specific organizational contexts, and difficulties in operationalizing theoretical concepts into actionable cybersecurity practices (NIST, 2018b). Furthermore, NIST (2018b) highlights the struggle agencies face in keeping pace with the rapid evolution of cyber threats while adhering to a framework that requires regular updates and adaptations. Bridging these gaps necessitates a concerted effort to enhance communication between policymakers, cybersecurity professionals, and on-the-ground IT staff, fostering a deeper understanding of RMF's principles and their practical implications. Moreover, developing continuous learning and adaptation strategies are crucial for ensuring that the theoretical underpinnings of RMF effectively translate into robust cybersecurity measures within federal agencies.

The integration of automation within RMF processes significantly enhances the efficiency and effectiveness of cybersecurity measures in organizations. Takamura et al. (2023) elucidates the numerous advantages that automation brings to RMF implementation, including streamlined compliance assessments, rapid identification, and mitigation of vulnerabilities, and the ability to maintain a continuous state of readiness against cyber threats. Automation not only reduces the manual labor associated with executing RMF tasks but also increases the accuracy and reliability of security controls and risk assessments. This advancement is pivotal in adapting to the dynamic nature of cybersecurity, where threats evolve at a pace that manual processes cannot match. By leveraging automation, organizations can achieve a more robust and resilient cybersecurity posture, ensuring that they can quickly respond to and mitigate potential risks as they arise.

Continuous monitoring plays a crucial role in enhancing cybersecurity measures, particularly in detecting and mitigating potential threats in real-time. Dempsey et al. (2020) emphasize the significance of continuous monitoring in safeguarding digital assets and sensitive information from cyberattacks. By implementing automated systems for monitoring network activities, organizations can effectively identify anomalous behaviors and unauthorized access attempts promptly. This proactive approach allows for timely responses and interventions, minimizing the impact of security breaches and enhancing overall resilience against evolving cyber threats. Automated tools streamline the monitoring process, enabling security teams to focus on analysis and decision-making rather than routine tasks. Therefore, investing in continuous monitoring solutions is essential for maintaining robust cybersecurity posture in today's dynamic threat landscape.

## VI. Case Studies

In the realm of cybersecurity within federal agencies, the RMF plays a pivotal role in enhancing security protocols and compliance measures. A comprehensive in-depth analysis of specific federal agencies' experiences with RMF reveals both the successes achieved and challenges encountered in its implementation (NIST, 2018a; 2018b). These case studies underscore the RMF's adaptability and its critical function in establishing robust cybersecurity postures tailored to each agency's unique operational environment. Through these analyses, insights into best practices, as well as common obstacles such as integration complexities and resource constraints, emerge, offering valuable lessons for other agencies in their cybersecurity endeavors. Such case studies not only highlight the practical application of the RMF but also contribute to a broader understanding of its impact on the federal cybersecurity landscape.

The compilation of case studies on the RMF across various federal agencies has led to the identification of critical lessons learned and best practices in cybersecurity management. These insights are invaluable for agencies striving to enhance their cybersecurity posture and compliance with federal standards. Key lessons emphasize the importance of stakeholder engagement, the need for clear communication across departments, and the integration of RMF processes into the organizational culture (NIST, 2018a; 2018b). Best practices identified include the early and continuous involvement of IT and cybersecurity teams in the RMF process, tailored training programs for staff, and the adoption of automation tools to streamline RMF tasks (NIST, 2023). These findings not only contribute to a deeper understanding of the RMF's application in federal contexts but also serve as a guide for other organizations navigating similar cybersecurity challenges.

VII. Recommendations

To effectively navigate the complexities and overcome the challenges associated with the implementation of the RMF, several strategic approaches have been recommended by cybersecurity experts. These strategies emphasize the importance of organizational readiness, including the development of comprehensive training programs for staff involved in RMF processes (NIST, 2018a; Takamura et al., 2023). Furthermore, fostering a culture of continuous improvement and adaptability are crucial for integrating RMF into the existing cybersecurity practices of federal agencies (NIST, 2018b). Enhanced collaboration across departments and leveraging automation for routine tasks are also pivotal strategies that can streamline the RMF implementation process, thereby mitigating potential obstacles and facilitating a smoother transition to robust cybersecurity frameworks (Takamura et al., 2023). Implementing these strategies require a concerted effort from all levels of the organization, ensuring that RMF adoption leads to improved security and compliance outcomes.

As the RMF continues to evolve, identifying future directions for its enhancement and associated research is crucial for maintaining the cybersecurity resilience of federal agencies. A key area of focus should include the integration of emerging technologies, such as artificial intelligence and machine learning, to automate and optimize RMF processes (Takamura et al., 2023). Additionally, research should explore the development of more adaptive RMF models that can swiftly respond to new and evolving cyber threats, ensuring that cybersecurity measures are both proactive and dynamic (NIST, 2018b). Further investigation into the human factors affecting RMF implementation, such as organizational culture and training, can also provide insights into enhancing the framework's efficacy (NIST, 2018a). These future directions not only promise to strengthen the RMF but also offer a pathway for continuous improvement in cybersecurity practices across federal agencies.

## VIII. Conclusion

The implementation of the RMF within federal agencies has revealed both significant advancements in cybersecurity practices and notable challenges that necessitate ongoing attention. Key findings underscore the RMF's effectiveness in establishing a structured and comprehensive approach to cybersecurity risk management, particularly through its emphasis on continuous monitoring and the integration of security controls (NIST, 2018a). However, challenges such as the complexity of RMF processes, the need for substantial training, and the integration of RMF into existing organizational cultures have been identified as barriers to its full effectiveness (NIST, 2018b; Takamura et al., 2023). These findings underscore the dual nature of RMF implementation. While it provides a robust framework for enhancing cybersecurity postures, its practical application requires careful navigation of organizational and procedural hurdles. Addressing these challenges through targeted strategies, such as increased automation and tailored training programs, remains essential for maximizing the RMF's potential benefits across federal agencies.

The RMF plays a pivotal role in achieving cybersecurity compliance within executive federal agencies, underpinning the essential strategies for protecting national security and sensitive information. The significance of RMF lies in its comprehensive approach to identifying, assessing, and managing cybersecurity risks, thereby ensuring that federal agencies can operate within a secure and resilient digital environment (NIST, 2018b). As highlighted by NIST (2018a) and further emphasized in subsequent studies (Takamura et al., 2023), the RMF not only facilitates compliance with federal cybersecurity standards but also promotes a culture of continuous improvement and adaptation to emerging threats. This framework's importance extends beyond mere compliance, serving as a cornerstone for fostering a proactive cybersecurity posture that is vital for safeguarding the nation's digital infrastructure.